\documentclass[sigplan,10pt]{acmart}

\startPage{1}

\setcopyright{none}

\usepackage{booktabs}   
\usepackage{subcaption} 

\begin{document}

\title[Towards A Systems Approach To Distributed Programming]{Towards A Systems Approach To Distributed Programming}


\author{Christopher S. Meiklejohn}
\orcid{nnnn-nnnn-nnnn-nnnn}             
\affiliation{
  \position{}
  \department{}             
  \institution{Universit\'e catholique de Louvain}           
  \streetaddress{}
  \city{Louvain-la-Neuve}
  \state{}
  \postcode{}
  \country{Belgium}                   
}
\affiliation{
  \position{}
  \department{}             
  \institution{Instituto Superior T\'ecnico}           
  \streetaddress{}
  \city{Lisbon}
  \state{}
  \postcode{}
  \country{Portugal}                   
}
\email{christopher.meiklejohn@uclouvain.be}         

\author{Peter Van Roy}
\orcid{nnnn-nnnn-nnnn-nnnn}             
\affiliation{
  \position{}
  \department{}              
  \institution{Universit\'e catholique de Louvain}            
  \streetaddress{}
  \city{Louvain-la-Neuve}
  \state{}
  \postcode{}
  \country{Belgium}                    
}
\email{peter.vanroy@uclouvain.be}          

\begin{abstract}
It is undeniable that most developers today are building distributed applications.  However, most of these applications are developed by composing existing systems together through unspecified APIs exposed to the application developer.  Systems are not going away: they solve a particular problem and most applications today need to rely on several of these systems working in concert.  Given this, we propose a research direction where higher-level languages with well defined semantics target underlying systems infrastructure as a middle-ground.
\end{abstract}

\ccsdesc[500]{Software and its engineering~General programming languages}
\ccsdesc[300]{Social and professional topics~History of programming languages}


\maketitle

\section{Distributed Programming}

Applications today are inherently distributed.  Even if you are requesting a ride through a popular ride sharing service such Uber or Lyft, your request is being handled by several microservices running in the data center, with state replicated and stored across several different databases.~\cite{uber}

These different systems and databases each make different guarantees to the application developer and each has its own semantics.  This puts additional burden on the application developer; not only does she need to implement the business logic required to build the application, she also must ensure that the composition of systems being used is correct and preserves application invariants.  To provide a concrete example, Uber's ride matchmaking service involves three microservices for matching supply to demand, where data is stored in both durable storage and message queues for workflow management of the ride.

Systems composition and management of data consistency across multiple systems is a difficult challenge.  Not only do these systems provide different guarantees through their APIs (consider the case of composing a system providing at-least-once event delivery with a system that requires at-most-once delivery of events), these APIs are largely defined by their implementation with no formal semantics or other way to guarantee application correctness.  In one example discovered by Kingsbury~\cite{kafka-kingsbury} and later formalized by Alvaro et al.~\cite{alvaro2015lineage}, the Apache Kafka system, when managed by Apache Zookeeper, can acknowledge writes as durable and later lose the writes because of a incorrect interaction between the two systems.  In this example, it is important to highlight that each of these systems are believed to operate correctly in isolation, but these guarantees do not extend to the composition of these systems.

Historically, there have been two approaches taken to solve the challenge of distributed programming: greenfield language and runtime development, and work on retrofitting existing systems for distribution, each of which has had little widespread success.

In terms of greenfield language development, the Argus~\cite{liskov1988distributed} and Emerald~\cite{raj1991emerald} systems attempted to provide new languages and runtime systems for distributed application development.  These systems provided features that aided developers building distributed applications: namely, serializable transactions, support for asynchronous programming, and process/object mobility.  However, greenfield language and runtime development is difficult from an adoption point-of-view: application developers want to work with languages with an established community, a proven runtime system and efficient tooling.

In terms of retrofitting existing languages and systems, one notable example is CORBA.  CORBA attempted to solve the problem of distributed programming by allowing objects (in an object oriented programming language) to live anywhere on a network of machines, each using different languages and system architectures.  CORBA would take care of object migration, serialization, and made remote calls transparent to the application developer: they appeared synchronous and local.  While successful in aiding programmers who desired to write simple distributed applications, scaling these applications and dealing with the realities of distributed programming: namely, latency, partial failure, and concurrency, during execution was extremely challenging given that distribution was transparent to the application developer.~\cite{waldo1997note}  

A clear tension exists between these two extremes: languages and runtime systems that are designed for distribution will always be ideal, however unrealistic because developers want to build applications on proven systems, with proven languages.  Further exacerbating the issue, is the approach taken by the systems research community, where systems are developed in isolation to solve a particular problem; many of these systems have historically been industrialized (databases, queueing systems, etc.) and therefore exist as isolated components in a larger composed system.  These systems typically have no formal semantics, and have been only empirically validated and not formally verified.  

We believe that a promising direction for the programming languages community is to try to solve for the middle-ground.  Is it possible to treat existing systems as a backend to a general purpose compiler for distributed programming?  We believe so!  Our work on Lasp~\cite{meiklejohn2015lasp}, a restricted programming model for distributed programming is a first step towards this direction.    

\section{Lasp}

Lasp is a declarative, functional programming model for large-scale distributed computing that leverages replicated abstract data types, called Conflict-Free Replicated Data Types~\cite{shapiro2011conflict}, to ensure value convergence under concurrency using a merge function for any two copies of replicated state.  Lasp is implemented as a library in the Erlang programming language, allowing interoperability and composition with existing Erlang applications.  

Given Lasp is built assuming weak consistency, we can operate the Lasp system on a variety of different underlying infrastructures.  

We highlight some of the properties of Lasp below.

\paragraph{Specialization.}  There exists several implementations for each type of CRDT, and the Lasp system has the ability to specialize the implementation at both compile time and runtime; for instance, if your application never needs to remove an item from a collection, the implementation can be specialized to a CRDT set that does not model removals, which is more efficient in space.  Right now, this is a manual process, but we believe that this should be able to be mechanized with the use of an effects system.

\paragraph{Data storage.}  The Lasp system relies on an underlying data store for storage of the CRDTs: this underlying storage does not need to provide a particular level of consistency, nor replication, because the programming model and data replication layers live above the underlying store.  Lasp supports both built-in Erlang data stores, and has been extended to use both the Riak distributed data store and the Redis data store.

\paragraph{Network topology agnostic.}  Determining the network topology that the system will run on is a runtime parameter: no application code has to be changed to alter the communication paths between nodes.  In our current version, Lasp applications can run in either client/server, full mesh, or in peer-to-peer mode, all specified at runtime.  This is configurable through an external membership service called by the runtime, and could easily be integrated with a system like Apache Zookeeper, if one desired.

\paragraph{Configurable synchronization.}  Lasp applications are written using shared state.  Again, an option that is configurable at runtime, is how often nodes in the system should propagate their state to other nodes in the system.  The system provides the option to propagate changes immediately to all nodes in the system, propagate every $N$ changes, or propagate based on a timer interval: these settings do not alter program behavior, but only alter when changes become visible to other nodes in the system. 

\section{Moving Forward}

We believe that the success of distributed computing relies on tighter integration between the underlying infrastructure and application code.  However, the majority of research today on distributed computing is focused in the database and systems communities, where the focus is on building standalone systems for solving individual problems.  While this direction has been incredibly fruitful, application developers typically need many of these systems working in concert to solve an actual business requirement.  Therefore, application developers devote a significant amount of effort to composing systems together using APIs with underspecified semantics, hoping for the best.  We believe that the programming language community can make a significant impact here by applying principled techniques to building restricted programming models for distributed computing that leverage infrastructure being created by the systems community.



\bibliography{obt-2018}
%
%
%

\end{document}